\documentclass[preprint,superscriptaddress]{revtex4}
\usepackage{graphics,epsfig}
\usepackage{epstopdf}
\usepackage{graphicx}
\usepackage{dcolumn}
\usepackage{amsmath}
\begin{document}
\title{Hayward black holes in Einstein-Gauss-Bonnet gravity}

\author{Arun Kumar}
\email{arunbidhan@gmail.com}
\affiliation{Centre for Theoretical Physics,
Jamia Millia Islamia, New Delhi 110025,
India}

\author{Dharm Veer Singh}
\email{veerdsingh@gmail.com}
\affiliation{Department of Physics, Institute of Applied Science and Humanities,
G. L. A. University, Mathura, 281406 India.}

\author{Sushant G. Ghosh}
\email{sgghosh@gmail.com}
\affiliation{Centre for Theoretical Physics,
Jamia Millia Islamia, New Delhi 110025,
India}
\affiliation{Multidisciplinary Centre for Advanced Research and Studies (MCARS),\\ Jamia Millia Islamia, New Delhi 110025, India}
\affiliation{Astrophysics and Cosmology Research Unit,
School of Mathematics, Statistics and Computer Science,
University of KwaZulu-Natal, Private Bag X54001,
Durban 4000, South Africa}

\begin{abstract}
The Hayward metric is a spherically symmetric charged regular black holes, a modification of the Reisnner-Nordstr$\ddot{o}$m black holes of Einstein's equations coupled to nonlinear electrodynamics. We consider Einstein-Gauss-Bonnet gravity (EGB) coupled to nonlinear electrodynamics to present an exact five dimension ($5D$) Hayward black holes with a regular center, having inner (Cauchy) and outer (event) horizons which go over to Boulware-Desser black holes when the charge is switched off ($e=0$). The presence of charge $e$ leads the modification in thermodynamical quantities, and it has also been shown that the Hawking-Page like phase transition can be achieved. The specific heat shows divergence at the horizon radius $r=r_C$ (critical radius), where the temperature has a maximum. Our result in the limit, $e\to0$, reduces vis-a-vis to the $5D$ Boulware-Desser solutions.
\end{abstract} 

\maketitle

\section{\label{sec:level1}Introduction}
 
 Black holes are the exact solution of the Einstein's general relativity, appear to exist in the universe,  with singularities form inside them \cite{1}.   However, the existence of a singularity means spacetime ceases to exist signaling the breakdown of general relativity, requiring modifications that believably include quantum theory. One of the steps in this direction, regular (i.e. non-singular) black holes have widely considered resolving the singularity problems, dating back to Bardeen \cite{Bardeen:1968} who gave first regular black hole model by Bardeen \cite{Bardeen:1968}, according to whom there are horizons but there is no singularity.  The Bardeen's model was motivated by the idea of Sakharov \cite{Sakharov:1966} who suggested a de Sitter core with equation of state $P=-\rho$ or $T_{ab}=\Lambda g_{ab}$ to get a  regular model without singularities, which could provide a proper discrimination at the final stage of gravitational collapse, replacing the future singularity \cite{Gliner:1966}.
 One can find Bardeen's metrics which is spherically symmetric, static, asymptotically flat, have a regular center, and for which the sources are physically reasonable, with causal structure is similar to that of a Reissner-Nordstr¨om black hole, satisfying the weak energy condition and is also an exact solution \cite{AGB,AGB1,ABG99}. Later, there  has been significant attempts for regular black hole models \cite{Ansoldi:2008jw,Lemos:2011dq,Zaslavskii:2009kp,Bronnikov:2000vy}, and more recently \cite{Xiang,hc,lbev,Balart:2014cga,singh,fr1,dvs,ak}. However, these black hole models are concentrated on Bardeen's idea and they have similar properties. The generalization of these stationary
regular black holes to the axially symmetric case, Kerr-like black holes, was addressed recently via Newman Janis algorithm \cite{Bambi,Ghosh:2014pba}, and thereafter more rotating regular black holes were proposed \cite{Neves:2014aba,Toshmatov:2014nya,Ghosh:2014hea,Larranaga:2014uca}. 

Hayward \cite{hayward} proposed, Bardeen-like, regular space-times are given that describe the formation of a black hole from an initial vacuum region which has a  finite density and pressures, vanishing rapidly at large small and behaving as a cosmological constant at a small distance.    It is a simple exact model of general relativity coupled to electrodynamics and hence Hayward black hole has attracted significant attention in various studies, like  Quasinormal  modes  of the black holes Lin by {\it et al} \cite{Lin:2013ofa}, The geodesic equation of a particle by Chiba and Kimura \cite{Chiba:2017nml}, wormholes from the regular black hole \cite{Kuhfittig:2013jna,Halilsoy:2013iza} with  their stability  \cite{Sharif:2016wvq}, black hole thermodynamics \cite{mn} and related properties \cite{Mehdipour:2016vxh, Perez-Roman:2018hfy,Abdujabbarov:2016hnw}, and strong deflection lensing \cite{Zhao:2017cwk}.  The rotating regular Hayward's metric has been studied as a particle accelerator \cite{Gwak:2017zwm, Amir:2016nti, Amir:2015pja}.  

In the last few decades, there has been a noteworthy number of attempts in higher dimensions gravity in order to understand the low-energy limit of string theory. The Einstein-Gauss-Bonnet gravity is a very important higher dimensional generalization of Einstein’s gravity which was suggested by Lanczos \cite{5}, and then rediscovered by David Lovelock \cite{6}. The study of Einstein-Gauss-Bonnet theory becomes very important since it provides a broader set up to explore a lot of conceptual issues related to gravity. This theory is completely free of ghost and the order of the field equations in the Einstein-Gauss-Bonnet theory is no higher than two. Since their outset, there has been a lot of attempts to obtain the black hole solution, but  Boulware and Deser were the first who obtain the exact black hole solution in the Einstein-Gauss-Bonnet gravity \cite{7,8}. After that several number of exact black hole solutions with their thermodynamical properties has been discussed by various authors \cite{9,9.1,9.2,9.3,9.4,9.5,9.6,9.7,9.8,9.9,9.10,9.11,10,11,12,13,13.1,14}. Several black hole solutions with matter source generalizing the Boulware-Desser solution have also been explored \cite{15,18,18.1,19}.

A natural question to ask: what’s the effect of the Einstein-Gauss-Bonnet correction on the regular black holes and their properties? In order to answer this question, one would first need a regular solution for the Einstein-Gauss-Bonnet theory. It is the purpose of this paper to obtain a $5D$ spherically symmetric and static Hayward-like black holes solution of the Einstein–Gauss–Bonnet gravity. It turns out that the metric purposed here is an exact black hole model of Einstein-Gauss-Bonnet having minimal coupling with nonlinear electrodynamics thereby it is the generalization of the Boulware-Desser solution. In turn, we analyze the thermodynamical properties of these models and find that black holes have a stable remnant, and the thermodynamic stability of these models also has been analyzed. We show that Hawking-Page transition possible, but the relation between entropy and horizon is no longer valid.

The paper is organized as follows, we obtain a Hayward-like metric in $5$ dimensions with a regular center and also give the relevant field equations of EGB gravity minimally coupled to nonlinear electrodynamics. The horizon structure of the $5D$ EGB-Hayward black holes metric has also been investigated. In Sec. III, we do black hole mechanics analysis of $5D$ EGB-Hayward models. The stability and black hole remnant are also discussed. The article has been ended with concluding remarks in Sec. IV. We shall adopt the signature $(-,+,+,+,+)$ for metric and use the units $8\pi G=c=1$. 
\section{$5D$ Exact Hayward-like Black Holes in EGB gravity}
The interest in the EGB theory arose mainly because it appears as the low energy limit of string theory \cite{7}. The general relativity with minimal  coupling with nonlinear electrodynamics leads to exact spherically symmetric regular black holes \cite{AGB,AGB1,ABG99,dym1,dym2}. The two most famous exact black holes are Bardeen \cite{Bardeen:1968} and Hayward \cite{hayward} regular black holes. Here, we are interested in  the Hayward-like black hole solution with a regular center in the EGB gravity in $5D$ spacetimes. The simplest action of EGB theory minimally coupled to nonlinear electrodynamics reads
\begin{equation}
\label{action} 
\begin{split}
S=\frac{1}{k_5}\int
d^5x\sqrt{-g}\left[R+\alpha\,( {R^2-4R_{ab}R^{ab}+R_{ab c d}R^{ab c d})-\mathcal{L}(F)}
\right],
\end{split}
\end{equation}
where $R$, $R_{ab}$, are Ricci scalar and tensor respectively and $R_{abcd}$ is Riemann tensor and $k_5=16\pi G_5$. The $\alpha\geq0$ is Gauss-Bonnet coupling constant having the dimensions of $(length)^2$. One recovers $5D$ Einstein gravity in low energy limit for small curvatures, varying action (\ref{action}), we can get following field equations \cite{3, ghosh8,ghosh14}:
\begin{eqnarray}
&&{G}_{a b} +\alpha {H}_{a b} = T_{ab},
\label{egb1}
\end{eqnarray}
\begin{eqnarray}
 \nabla_{a}\left(\frac{\partial \mathcal{L}(F)}{\partial F}F^{a b}\right)=0\quad \text{and} \quad \nabla_{\mu}(* F^{ab})=0,
 \label{egb3}
\end{eqnarray}
where
\begin{eqnarray}
&&G_{ab}=R_{ab}-\frac{1}{2}g_{ab}R,\nonumber\\
&&{H}_{ab}=2\Bigl[RR_{ab}-2R_{a c}R^{c}_{b}-2R^{c d
}R_{a c bd} +R_{a}^{~c d e}R_{b c d e}\Bigr]-{1\over 2}g_{ab}{L}_{GB}
\end{eqnarray}
Remarkably, the equations of motion (2) do not have the derivatives of a metric function of order higher than two which means the theory does not suffer from the ghost \cite{Ghosh:2016ddh}.
\begin{eqnarray}
&&T_{ab}=2\left[\frac{\partial \mathcal{L}(F)}{\partial F}F_{a c}F_{b}^{c}-g_{a b}\mathcal{L}(F)\right].
\label{emt}
\end{eqnarray}

The function $\mathcal{L}(F)$ is an arbitrary function of $F=F_{ab}F^{ab}/4$ with $F_{ab}=\partial_{a}A_b-\partial_{b}A_a$ is electromagnetic field tensor, with, $\mathcal{L}(F) \approx F$ describes the linear Maxwell theory. For the Hayward-like regular black hole solution, the Lagrangian density in $5D$ spacetimes calculated as
\begin{equation}
\mathcal{L}(F)= \frac{3}{2 s e^2}\frac{({2 e^2 F})^{4/3}}{(1+(\sqrt{2e^2F})^{4/3})^2},
\label{nonl1}
\end{equation}
where $s$ is a positive constant. The field tensor $F_{ab}$ in $5D$ spacetime \cite{dvs18}
\begin{equation}
F_{ab}=2\delta^{\theta}_{[a}\delta^{\phi}_{b]}Z(r,\theta,\phi)=2\delta^{\theta}_{[a}\delta^{\phi}_{b]}e(r) \sin^2\theta\sin\phi.
\label{7}
\end{equation} 
with $F_{\theta\phi}, F_{\theta\psi}$ and $F_{\phi\psi}$ as the only non vanishing components. Eq (\ref{egb3}) gives $dF=0$ which, in turn implies $e(r)=e$ is a constant.
 Thus, the Field strength tensor simplifies to
\begin{equation}
F_{\theta\phi}=\frac{e}{r}\sin\theta,\qquad \text{and} \qquad  F=\frac{e^4}{2r^6}
\label{8}
\end{equation} and
\begin{eqnarray}\label{lf1}
\mathcal{L}(F)&=&\frac{3e^6}{2s(r^4+e^4)^2}
\end{eqnarray}
On using Eq. (\ref{lf1}) into Eq. (\ref{emt}), we obtain
\begin{equation}
T^t_t=T^r_r=\rho(r)= \frac{3 e^6}{s(r^4+e^4)^{2}}.
\end{equation}
The other components of energy momentum tensor are obtained by using the Bianchi identities and, they are given by
\begin{eqnarray}
&&T^{\theta}_{\theta}=T^{\phi}_{\phi}=T^{\psi}_{\psi}=\rho(r)+\frac{r}{3}\partial_r\rho(r).
\end{eqnarray}
Thus, the EMT is completely determined.
  To obtain a $5D$ static, spherically symmetric solutions  of Eq. (\ref{egb1}), we use the metric \textit{anstaz} \cite{Ghosh}
\begin{equation}
ds^2=-f(r)dt^2 +\frac{1}{f(r)}dr^2+r^2d\Omega_3^2,
\label{m1}
\end{equation}
where $d\Omega_3^2=d\theta^2+\sin^2\theta \,(d\phi^2+\sin^2\phi \,d\psi^2)$ is the metric  in the $3D$  hypersurface with volume $V_3$ and $f(r)$ is the metric function to be determined. On using equation (\ref{m1}), by solving Field equations (\ref{egb1}) we obtain the equations of motion
\begin{eqnarray}
&&f'-\frac{2}{r}(1-f)+\frac{4\alpha}{r^2}(1-f)f=\frac{3 e^6 }{s(r^4+e^4)^{2}},
\label{eom2}
\end{eqnarray}
\begin{eqnarray}
&&f''+\frac{4}{r}f'+\frac{2}{r^2}(1-f)+\frac{4\alpha}{r^2}\left[f''(1-f)+f'^2\right]=\frac{e^6(3e^4-5r^4)}{s(r^4+e^4)^3}.
\end{eqnarray} 
The Eq. (\ref{eom2}) admits an exact solution
\begin{equation}
  f(r)=1+\frac{r^2}{4\alpha}\left(1\pm\sqrt{1+\frac{8\alpha m  }{r^4+e^4}}\,\right).
\label{eqn:f}
\end{equation}
 Here, $m$ is a constant of integration having the relation with the Arnowitt-Deser-Misner ($ADM$) mass $M$ of the black hole via 
\begin{equation}
M=\frac{3V_3}{k_5}m.
\end{equation} 
   Here, $V_3$ is the volume of a 3-dimensional unit sphere. It is easy to show that the other field equations are also satisfied. Thus, (\ref{m1}) with metric function (\ref{eqn:f}) is an exact solution of EGB coupled to NED which encompasses the $5D$ Boulware-Desser solution \cite{7} as special case when $e=0$. The negative branch of the above solution is physical as it yields the $5D$ Hayward black holes, in the limit $\alpha\to0$, given by metric (\ref{m1}) with 
\begin{equation}
  f(r)\sim 1-\frac{mr^2}{r^4+e^4}.
\end{equation} 
and $5D$ Schwarzschild-Tangherlini black hole \cite{st} when the charge is also switched off ($e=0$). There is no such limit for the positive branch of solution. Henceforth, we shall call the solution (\ref{m1}) with metric function (\ref{eqn:f}) as $5D$ EGB-Hayward black holes. The flatness at centre, $r\to0$, requires that the  metric function (\ref{eqn:f}) behaves as

\begin{equation}
f(r)\sim 1+\frac{r^2}{l^2}
\end{equation}    
with $1/l^2=\left(1-\sqrt{1+{8\alpha m}/{e^4}}\right)/{4\alpha}$. Thus, the $5D$ EGB-Hayward solution has a central de-Sitter core. For large r
\begin{equation}
f(r)\sim 1-\frac{mr^2}{r^4+e^4}+\alpha \mathcal{O}(r^4+e^4)^{-2}
\end{equation} if we further switch off the magnetic charge $(e=0)$, metric function take the form \cite{9}
\begin{equation}
f(r)\sim 1-\frac{m}{r^2}+\alpha \mathcal{O}(r^{-4})
\end{equation} 

The regularity of $5D$ EGB-Hayward black holes can be addressed by calculating invariants, Ricci scalar $R$, Ricci square $R_{ab}R^{ab}$ and Kretschmann  scalar $R_{abcd}R^{abcd}$, which are calculated as
\begin{eqnarray}
&&\lim_{r\to 0} R =\frac{5}{\alpha}\left[-1+\left(1+\frac{8\alpha m}{e^{4}}\right)^{1/2}\right],\nonumber\\
&&\lim_{r\to 0}R_{ab}R^{ab}=\frac{10}{\alpha^2}\left[1+\frac{4\alpha m}{e^{4}}\left(1+\frac{8\alpha m}{e^{4}}-\right)^{1/2}\right],\nonumber\\
&&\lim_{r\to 0} R_{abcd}R^{abcd}=\frac{5}{\alpha^2}\left[1+\frac{4\alpha m}{e^{4}}-\left(1+\frac{8\alpha m}{e^{4}}\right)^{1/2}\right].\nonumber\\
\label{inv}
\end{eqnarray}
We found that, when $m \neq 0 \neq \alpha$, the invariants are well behaved everywhere including at $r=0$. Thus, the $5D$ Hayward like black holes has no singularity or they are regular.

 The weak energy condition states that $T_{ab}t^at^b\geq 0$ for all time like vectors $t^a$, ie., for any observer, the local energy density must not be negative. 
Hence, the energy conditions require $\rho\geq 0$ and $\rho+P_i\geq 0$,
\begin{eqnarray}\label{ec}
\rho=\frac{3 e^4M }{(r^4+e^4)^{2}}\nonumber\\
\rho+P_2=\rho+P_3=\rho+P_4=\frac{6e^8 M}{(r^4+e^4)^{3}}.
\end{eqnarray}
   Eq. \eqref{ec} signifies the validity of weak energy condition for $5D$ EGB-Hayward black holes.

Next, we study the structure of horizons of $5D$ EGB-Hayward black holes. It turns out that $g^{rr}=f(r_H)=0$ is only coordinate singularity implies the presence of horizons. Thus, the horizons are zeros of
\begin{equation}
              r_H^6+(2\alpha-m)r_H^4+e^4r_H^2+2\alpha e^4=0.
 \label{rh}
\end{equation}
The location of horizon are real roots of Eq. (\ref{rh}).
\begin{eqnarray}
              r_+^2&=&\frac{1}{3}\left[m-2\alpha-\frac{2^{\frac{2}{3}}(3e^4-(m-2\alpha)^2)+\beta^2}{2^{\frac{1}{3}}\beta}\right]\\~~~~ \text{with}\nonumber\\ \beta&=&2(m-2\alpha)^2+12m\alpha^2-18(m+4\alpha)e^4\nonumber\\&&+\sqrt{\left(2(m-2\alpha)^2+12m\alpha^2-18(m+4\alpha)e^4\right)^2+4\left(3e^4-(m-2\alpha)\right)^3}
 \label{rh1}
\end{eqnarray}

\begin{figure}
\begin{tabular}{c c c c}
\includegraphics[width=.50\linewidth]{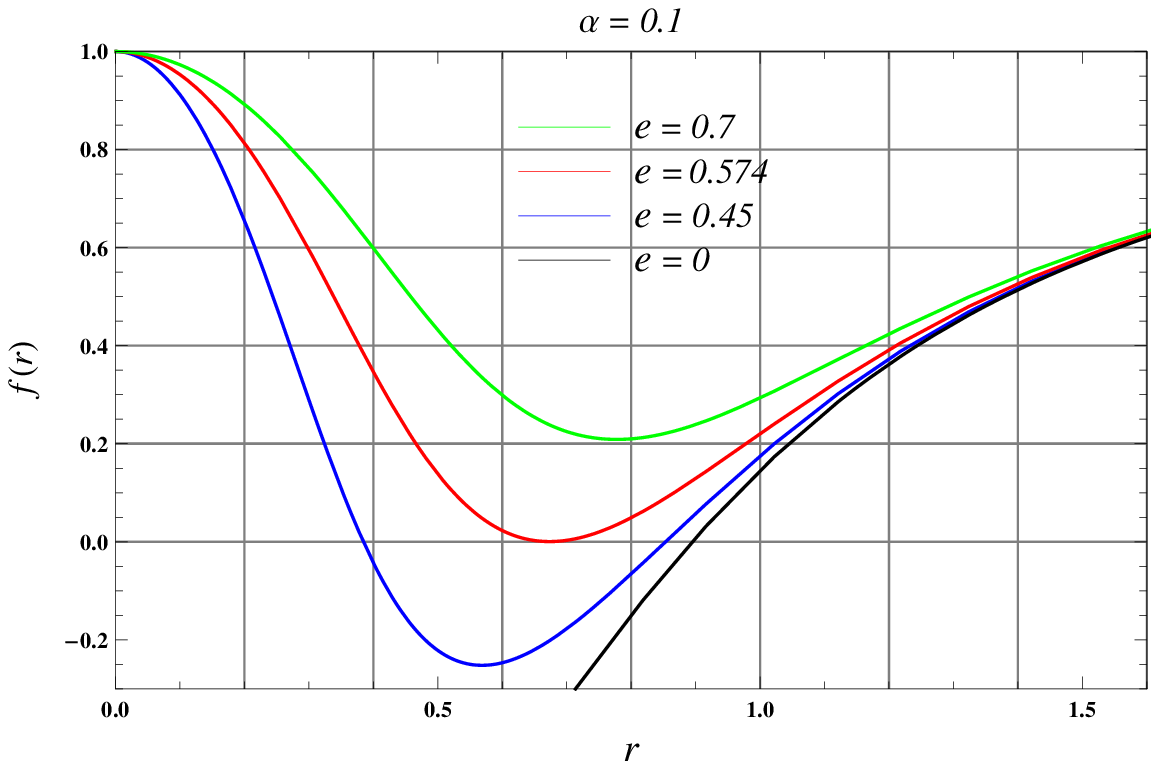}
\includegraphics[width=.50\linewidth]{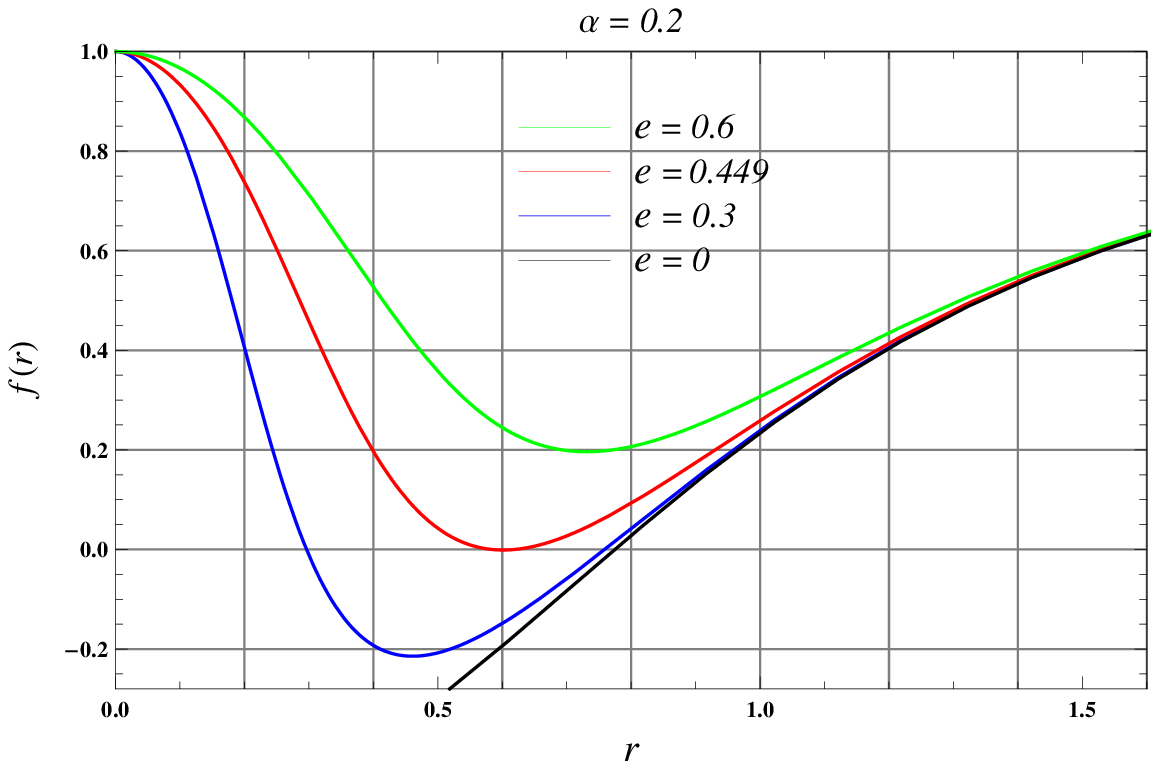}
\end{tabular}
\caption{The plot of $f(r)$  vs $r$ for  different values of charge  $e$, Gauss-Bonnet coupling constant $\alpha$.}
\label{fig:1}
\end{figure}

 We find that it's possible to find parameters $e$ and $\alpha$ such that Eq. (\ref{rh}) admits two positive roots $r_{\pm}$, with $r_-$ and $r_+$ are, respectively, representing the Cauchy and event horizons. By keeping, the value of mass $m$ and coupling constant $\alpha$  fixed, we come to find out that there exists a critical value of charge ($e_E$), such that the Cauchy ($r_-$) and the event horizons ($r_+$) coincide, i.e, $r_- = r_+$, corresponds to the extremal $5D$ EGB-Hayward black holes with degenerate horizon radius ($r_E=r_{\pm}$). So, when $e<e_E$, black holes with Cauchy and event horizons exist (cf. Fig. \ref{fig:1}) and for any value of charge $e>e_E$, there exist only a regular spacetime but not black holes. We also note that the size of the event horizon ($r_+$) decreases as we increase ($\alpha$) and increases with increase in the value of magnetic charge ($e$) as shown in Table \ref{tab:temp1}.
 \begin{center}
\begin{table*}[ht]
\begin{center}
\begin{tabular}{|l|l r l| l| l r l|}
\hline
\multicolumn{1}{|c|}{ }&\multicolumn{1}{c}{ }&\multicolumn{1}{c}{$\alpha=0.1$  }&\multicolumn{1}{c|}{ \,\,\,\,\,\, }&\multicolumn{1}{c}{ }&\multicolumn{1}{c}{}&\multicolumn{1}{c}{ $\alpha=0.2$ }&\multicolumn{1}{c|}{\,\,\,\,\,\,}\\
\hline
\multicolumn{1}{|c|}{ \it{e}} & \multicolumn{1}{c}{ $r_-$ } & \multicolumn{1}{c}{ $r_+$ }& \multicolumn{1}{c|}{$\delta$}&\multicolumn{1}{c|}{\,\,\it{ e}}& \multicolumn{1}{c}{$r_-$} &\multicolumn{1}{c}{$r_+$} &\multicolumn{1}{c|}{$\delta$}   \\
\hline
\,\,\,\,\,\,\,\,\,\,\,\, $0.1$\,\,& \,\,0.071\,\, &\,\,  0.894\,\,& \,\,0.823&\,\,\,\,\,\,\,\,\,\,\,\,\,\,\,0.1&\,\, 0.091\,\,&\,\,0.779\,\,&\,\,0.688\,\,
\\
\,\,\,\,\,\,\,\,\,\,\,\, $0.2$\,\,& \,\,0.146\,\, &\,\,  0.893\,\,& \,\,0.747&\,\,\,\,\,\,\,\,\,\,\,\,\,\,\,0.2&\,\, 0.187\,\,&\,\,0.771\,\,&\,\,0.584\,\,
\\
\,\,\,\,\,\,\,\,\,\,\,\, $0.3$\,\,& \,\,0.228\,\, &\,\,  0.887\,\,& \,\,0.659&\,\,\,\,\,\,\,\,\,\,\,\,\,\,\,0.3&\,\, 0.296\,\,&\,\,0.759\,\,&\,\,0.463\,\,
\\
\,\,\,\,\,\,\,\,\,\,\,\, $0.4$\,\,& \,\,0.326\,\, &\,\,  0.870\,\,& \,\,0.544&\,\,\,\,\,\,\,\,\,\,\,\,\,\,\,0.4&\,\, 0.439\,\,&\,\,0.714\,\,&\,\,0.275\,\,
\\
$e_E=0.574$\,\, &  \,\,0.673\,\,  &\,\,0.673\,\,&\,\,0\,\,& $e_E=0.449$&\,\, 0.601\,\,&\,\,0.601\,\,&\,\,0\,\,
 \\
 \hline
\end{tabular}
\end{center}
\caption{Inner horizon radius ($r_-$), the outer (event) horizon radius ($r_+$) and $\delta=r_+-r_-$ corresponding to the various different values of magnetic charge $e$.}
\label{tab:temp1}
\end{table*}
\end{center}

\section{Black hole Thermodynamics}
  Next, we calculate the thermodynamical quantities of EGB-Hayward black holes at the event horizon ($r_+$). By solving $f(r_+)=0$, we obtained the black hole mass 
\begin{equation}
M_+=\frac{3V_3r_+^2}{k_5} \left[\left(1 + \frac{2 \alpha}{r_{+}^2}\right)\left(1+\frac{e^{4}}{r_+^{4}}\right)\right].
\label{eqM}
\end{equation}
\noindent By switching off the magnetic charge ($e=0$) in Eq. (\ref{eqM}), one can obtain the EGB black hole mass \cite{9,cai02,ghosh14,Mayers88}
\begin{equation}
M_+=\frac{3V_3r_+^2}{k_5}[1+\frac{2\alpha}{r_+^2}]
\end{equation}
 and further for limiting case, ($e=0, \alpha \to 0$), Eq. (\ref{eqM}) reduces to $M_+=3V_3r_+^2/k_5$, which is the mass of $5D$ Schwarzschild-Tangherlini black hole \cite{ghosh8}. The Hawking temperature of the black hole is defined as $T=\kappa/2\pi$, with surface gravity ($\kappa$)
\begin{equation} 
\kappa= \left(-\frac{1}{2}\nabla_{a}\xi_{b}\nabla^{a}\xi^{b}\right)^{1/2}.
\label{temp21}
\end{equation}
 Using Eqs. (\ref{m1}) and (\ref{eqn:f}) in Eq. (\ref{temp21}), we obtain the temperature for the $5D$ EGB-Hayward black hole as
\begin{equation}
T_+=\frac{1}{4 \pi r_+} \left[\frac{2r_{+}^2-\frac{2e^{4}}{r_{+}^{4}}(r_{+}^2+4 \alpha)}{ (r_{+}^2+4\alpha)(1+\frac{e^{4}}{r_{+}^{4}})}\right].
\label{eqT}
\end{equation}

 The positive temperature $T_+>0$ requires $r_+^6>e^2(r_+^2+4\alpha)$. In the limit $e\to0$,  Eq. (\ref{eqT}) reduces to the Hawking temperature of EGB  black hole \cite{cai02,ghosh14,Mayers88}, 
\begin{equation}
T_+=\frac{1}{2\pi }\left(\frac{r_+}{r_+^2+4\alpha}\right),
\end{equation}
when $\alpha \to 0$, one recovers the temperature of $5D$ Hayward black hole,
\begin{equation}
T_+= \frac{1}{2\pi r_+}\left(\frac{r_+^4-e^4}{r_+^4+e^4}\right).
\label{eqT1}
\end{equation}
\begin{figure}[ht]
\begin{tabular}{c c c c}
\includegraphics[width=0.50\linewidth]{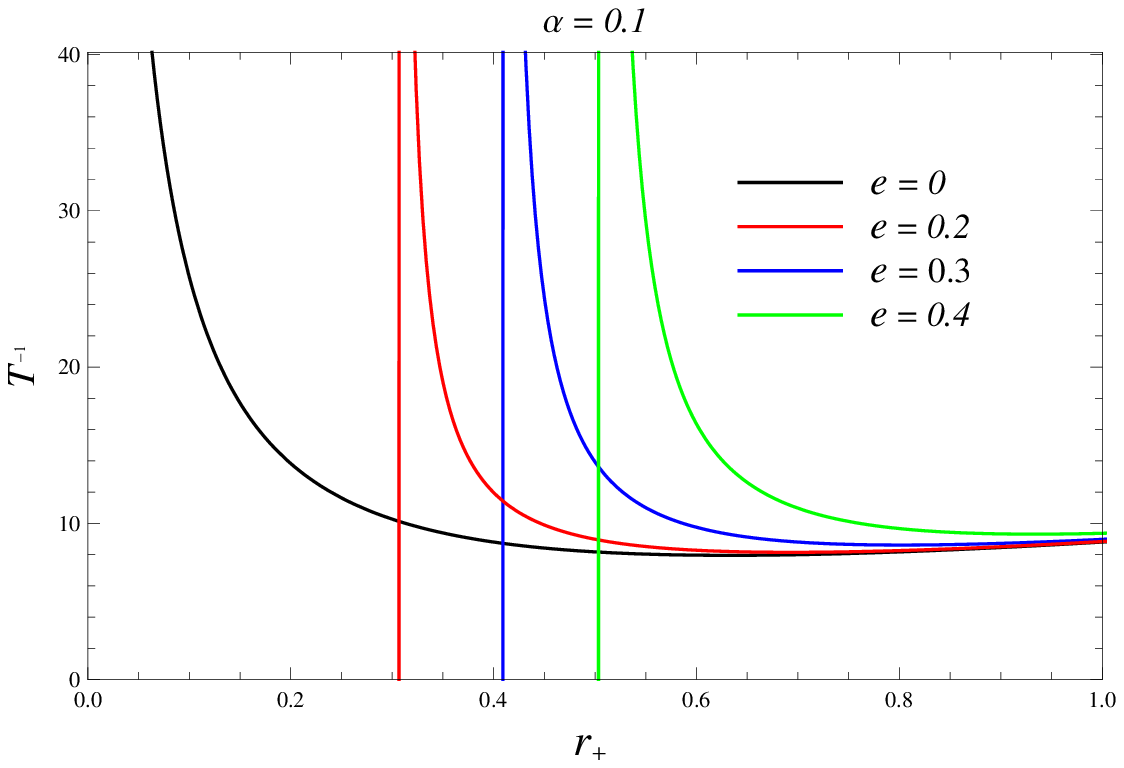}
\includegraphics[width=0.50\linewidth]{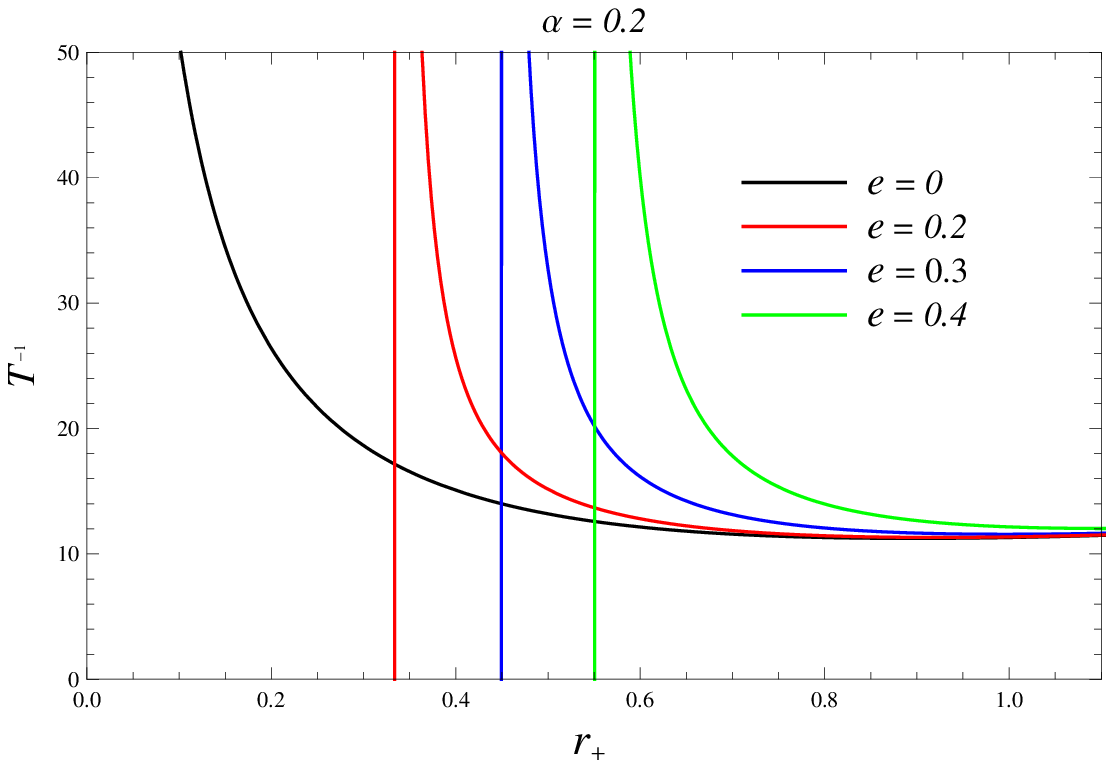}
\end{tabular}
\caption{ The inverse Hawking temperature ($T^{-1}_+$) vs horizon radius ($r_+$) for different values of $e$ and $\alpha$.}
\label{fig:th1}
\end{figure}

which further reduces to the temperature of the $5D$ Schwarzschild-Tangherlini black holes \cite{ghosh8}, $T_+=1/2\pi r_+$, when $e=0$. The behaviour of inverse Hawking temperature for different values of $e$ and $\alpha$ is depicted in Fig. \ref{fig:th1}. The inverse Hawking temperature decreases with an increase in the horizon radius. From Fig. \ref{fig:th1}, one can notice that the Hawking temperature of $5D$ EGB-Hayward black holes with a small horizon radius goes to zero, unlike the case of $5D$ Schwarzschild-Tangherlini black hole, for which temperature diverges at a smaller value of $r_+$. In Table \ref{tab:temp}, we have shown numerical values of maximum Hawking temperature $T_+^{Max}$ with corresponding radii $r_c^{T}$. It is noticed that the critical radius increases when we increase $e$ and $\alpha$ and the maximum Hawking temperature decreases for higher values of the critical radius.
\begin{center}
\begin{table*}[ht]
\begin{center}
\begin{tabular}{|l | l r l r | l r l r |r r| }

\hline
\multicolumn{1}{|c}{ }&\multicolumn{1}{c}{ }&\multicolumn{1}{c}{ }{$\alpha=0.1$  }{ \,\,\,\,\,\, }&\multicolumn{1}{c}{ }&\multicolumn{1}{c|}{  }&\multicolumn{1}{c}{ }&\multicolumn{1}{c}{$\alpha=0.2$}&\multicolumn{1}{c}{}&\multicolumn{1}{c|}{}\\
\hline
\multicolumn{1}{|c|}{ \it{e} } &\multicolumn{1}{c}{ 0 } &\multicolumn{1}{c}{ 0.4 } & \multicolumn{1}{c}{ 0.5 }& \multicolumn{1}{c|}{0.6} &\multicolumn{1}{c}{ 0}&\multicolumn{1}{c}{0.4} &\multicolumn{1}{c}{ 0.5}   & \multicolumn{1}{c|}{0.6} \\
\hline
\,\,\,\,\,\,\,$r_c^T$\, &\,\,0.614\,\,& 0.911\,\,\,\,\,\,\, &~~\,\,1.077\, &  ~~\,\,1.218\,\, &\,\,0.876\,\,&~~\,\,1.092\,\,&\,\,~~1.238\,\,&~~\,\,1.339\,\,
 \\
\,\,\,\,\,\,$T_+^{Max}$&\,\,0.125\,\,&\,\,  0.106\,\,\,\,\,\,\, &~~\,\,0.097\,& ~~\,\,  0.089\,\,&\,\,0.088\,\,&\,\,0.082\,\,& \,\,~~0.078\,\,&~~\,\, 0.073\,\,
\\
 \hline
\end{tabular}
\end{center}
\caption{The maximum black hole temperature ($T_+^{Max}$) and corresponding horizon radius ($r_c^{T}$) with various values of charge ($e$).}
\label{tab:temp}
\end{table*}
\end{center}
To construct the entropy of our black hole solution, we employ the first law of black hole thermodynamics  \cite{ghosh8}
\begin{equation}
dM_+=T_+\,dS_++\phi de,
\end{equation}
where $e$ is a magnetic charge and $\phi$ is its corresponding potential. One finds the following expression for the black hole entropy $S$ for constant magnetic charge $e$,
\begin{eqnarray}
S_+=\int\,T_+^{-1}\partial M_+=\int\,\frac{1}{T_+}\frac{\partial M_+}{\partial r_+}dr_+,
\label{eqS}
\end{eqnarray}
By plugging (\ref{eqM}) and (\ref{eqT}) into (\ref{eqS}), we get the following entropy expression for $5D$ EGB-Hayward black holes
\begin{equation}
S_+=4\pi\int\left(1+\frac{e^4}{r_+^4}\right)(r^2_++4\alpha) \,dr_+.
\end{equation}
which can be integrated exactly to
\begin{eqnarray}
S_+=\frac{4\pi V_3 r^3_+}{k_5}\left[1 + \frac{12\alpha}{r_+^2} -\frac{e^4}{r_+^6}(3r_+^2 +4\alpha)\right].
\label{entropy}
\end{eqnarray}
 Note that the last two factors in the parenthesis of Eq. (\ref{entropy}) modifies entropy and area law $S=A/4$ is no longer valid. When one take $e=0$, Eq. (\ref{entropy}) reduces exactly to the entropy of the EGB black hole \cite{cai02,ghosh14, Mayers88}
\begin{eqnarray}
S_+=\frac{4\pi V_3 r^3_+}{k_5}\left[1+\frac{12\alpha}{r_+^2}\right].
\label{entropy}
\end{eqnarray}
\begin{figure}[ht]
\begin{tabular}{c c c c}
\includegraphics[width=0.5\linewidth]{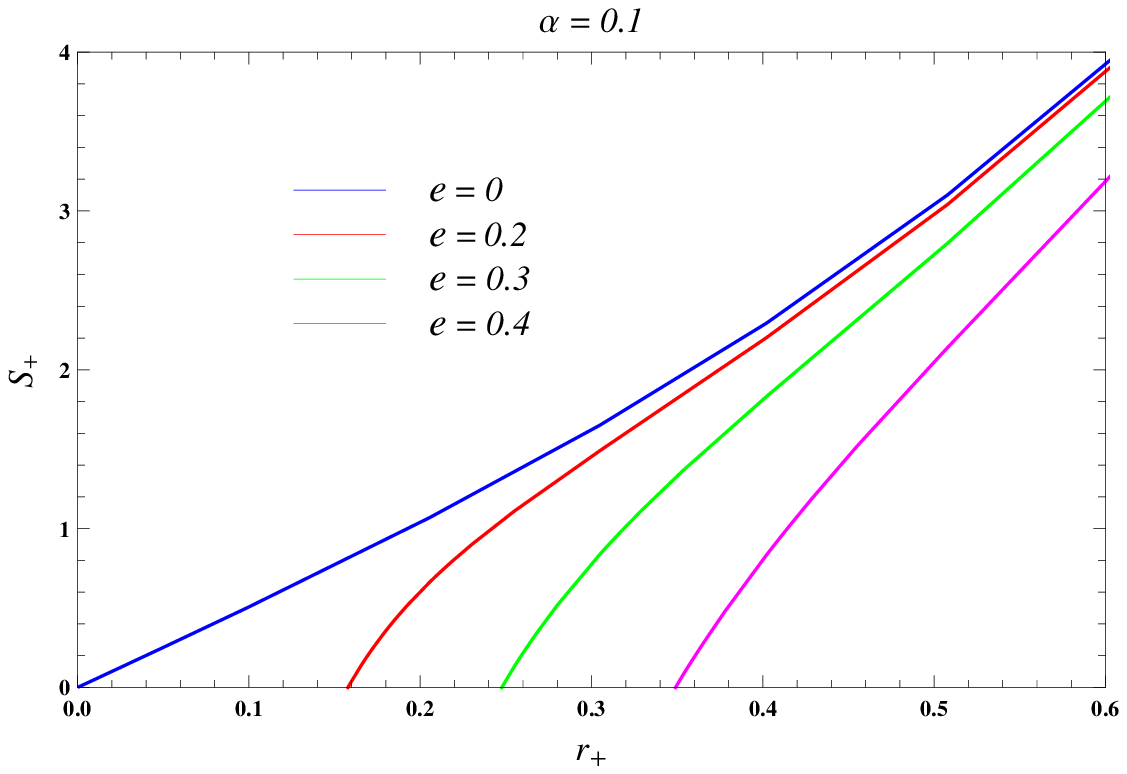}
\includegraphics[width=0.5\linewidth]{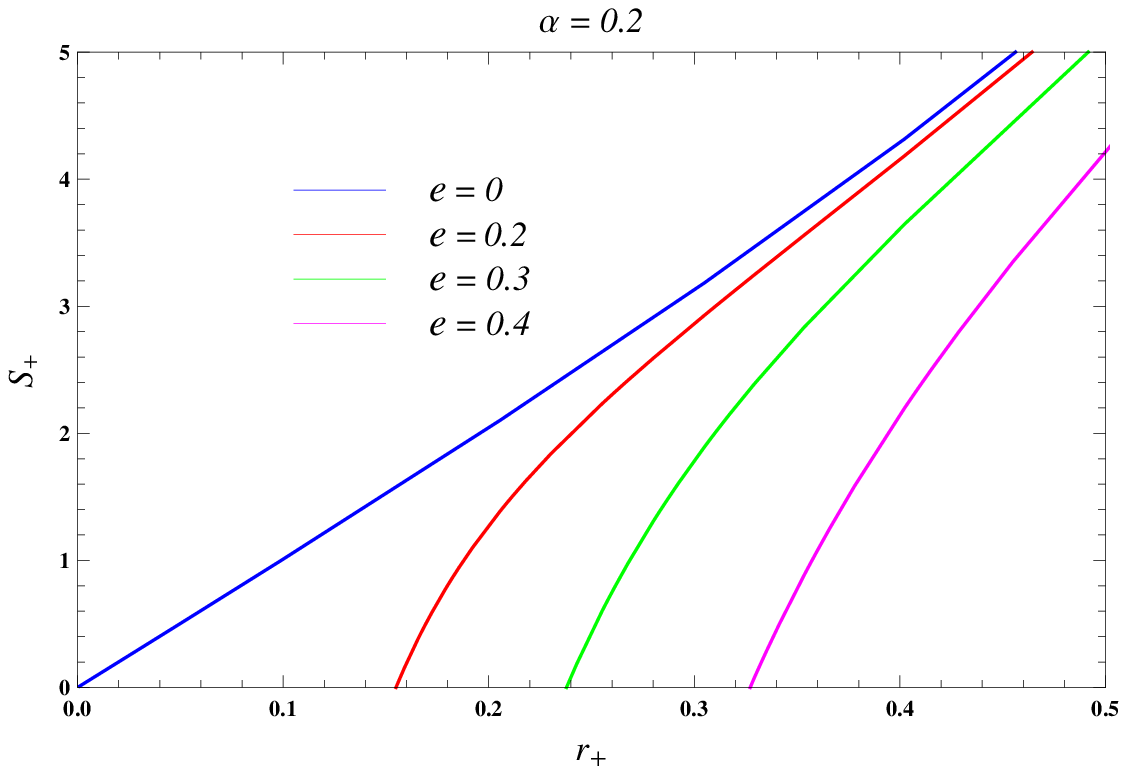}
\end{tabular}
\caption{Entropy $S_+$ \textit{vs} event horizon $r_+$ for different $e$ and $\alpha$.}
\label{entropy1}
\end{figure}
 Taking $\alpha \to 0$, we recover $S_+=4\pi V_3 r_+^3/k_5$, the entropy of $5D$ Schwarzschild-Tangherlini black hole \cite{ghosh8}, which obeys the area law. Thus, one can conclude that the presence of $\alpha$ and $e$, made the entropy area law invalid.

We study the behaviour of Gibb's free energy for the global stability of black hole thermodynamics. The expression of Gibb's free energy is expressed as \cite{Herscovich}
\begin{equation}
G_+ =M_+-T_+S_+,
\label{freef}
\end{equation}
substituting the value of Eq. (\ref{eqM}), Eq. (\ref{eqT}) and Eq.  (\ref{entropy}) in Eq. (\ref{freef}), we obtain
\begin{eqnarray}
G_+=&&-\frac{3V_3}{k_5}\left[\frac{\left[r_+^2(3r^8_++16e^4r_+^4+3e^8)+2\alpha(-3r_+^8-18e^4r_+^4+e^8)\right](r_+^2+4\alpha)-4r_+^8(e^4+r_+^4)}{3r_+^4(e^4+r_+^4)(r_+^2+4\alpha)}\right]\nonumber\\
\end{eqnarray}

\begin{figure}
\begin{tabular}{c c c c}
\includegraphics[width=0.5\linewidth]{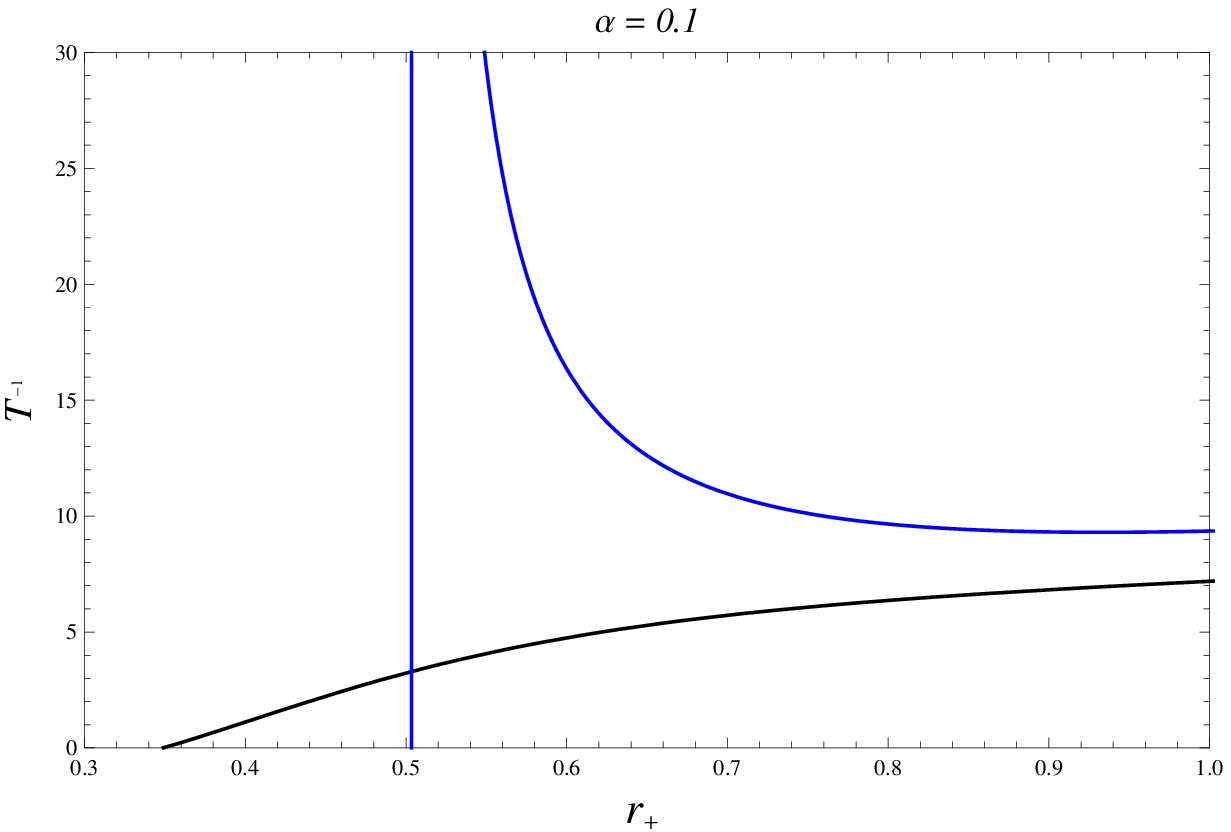}
\includegraphics[width=0.5\linewidth]{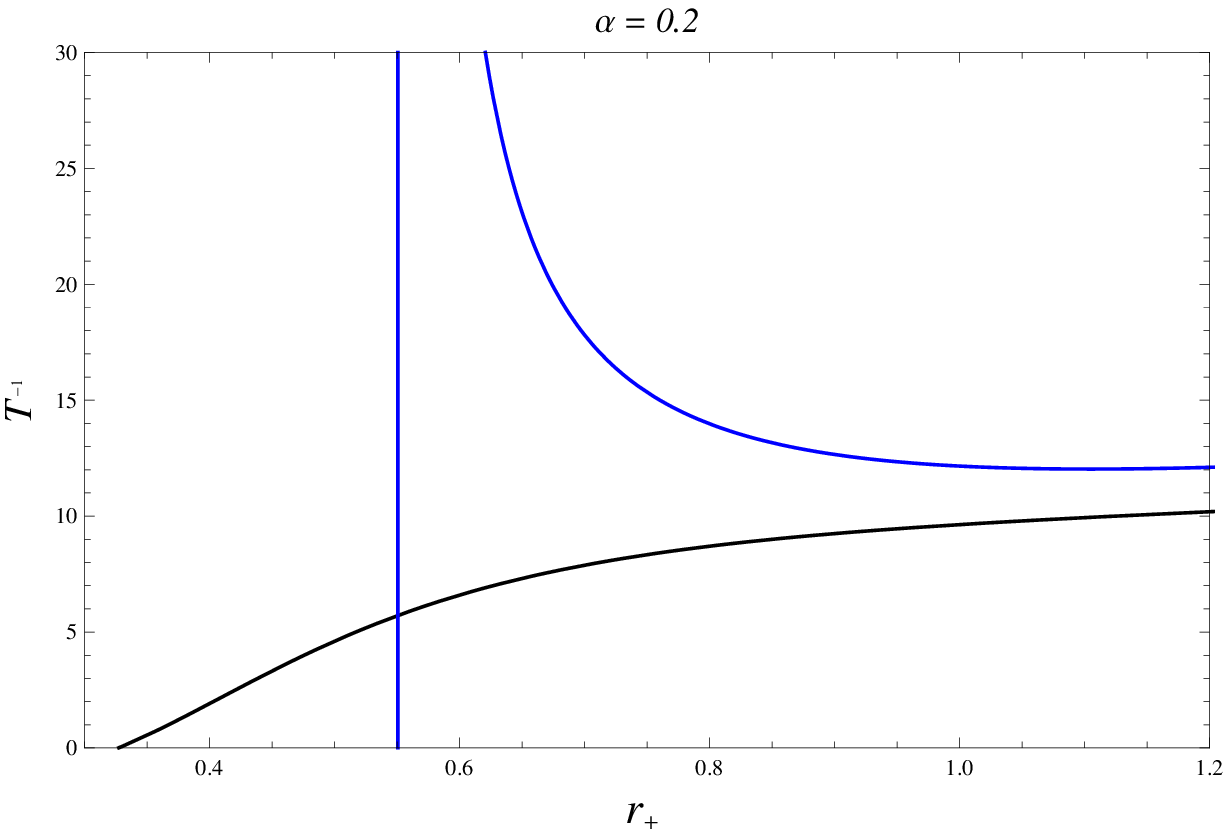}
\end{tabular}
\caption{ The inverse Hawking temperature \textit{vs} event horizon ($r_+$) for the case $e=0.5$ and $\alpha=0.1~~ \text{and}~~ 0.2$. The Black curve corresponds to $T_C^{-1}$ and the blue one to $T_+^{-1}$.}
\label{temp2}
\end{figure}

In the limit $\alpha \to 0$, gives the expression for the $5D$ Hayward black hole as
\begin{eqnarray}
G_+=&&-\frac{3V_3}{k_5}\left[\frac{2r_+^2(r_+^4-e^4)(r_+^4-3e^4)-(r_+^4+e^4)^2}{3r_+^2(r_+^4+e^4)}\right].
\end{eqnarray} In the limiting case, $e=0$, we obtain the free energy of EGB black holes \cite{9}
\begin{equation}
G_+=-\frac{V_3}{k_5}\left[\frac{-4r_+^4+3(r_+^2-3\alpha)(r_+^2+4\alpha)}{r_+^2+4\alpha}\right]
\end{equation} 
\begin{figure}[ht]
\begin{tabular}{c c c c}
\includegraphics[width=0.5\linewidth]{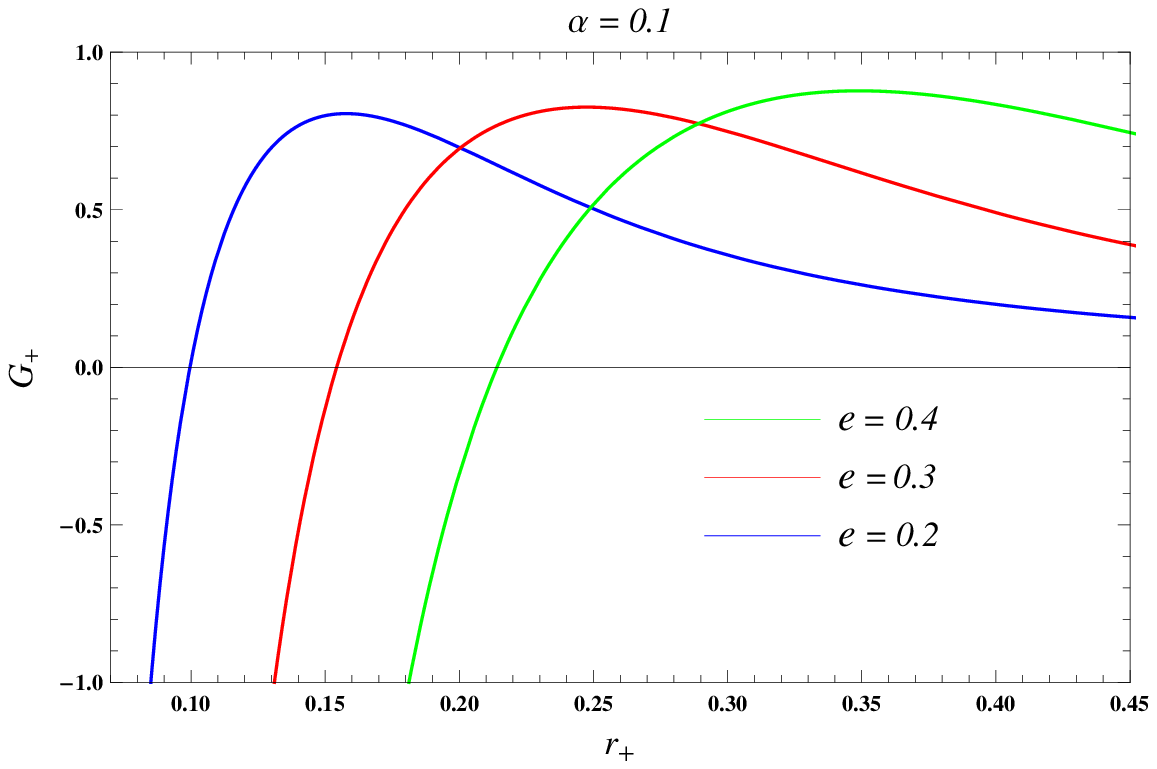}
\includegraphics[width=0.5\linewidth]{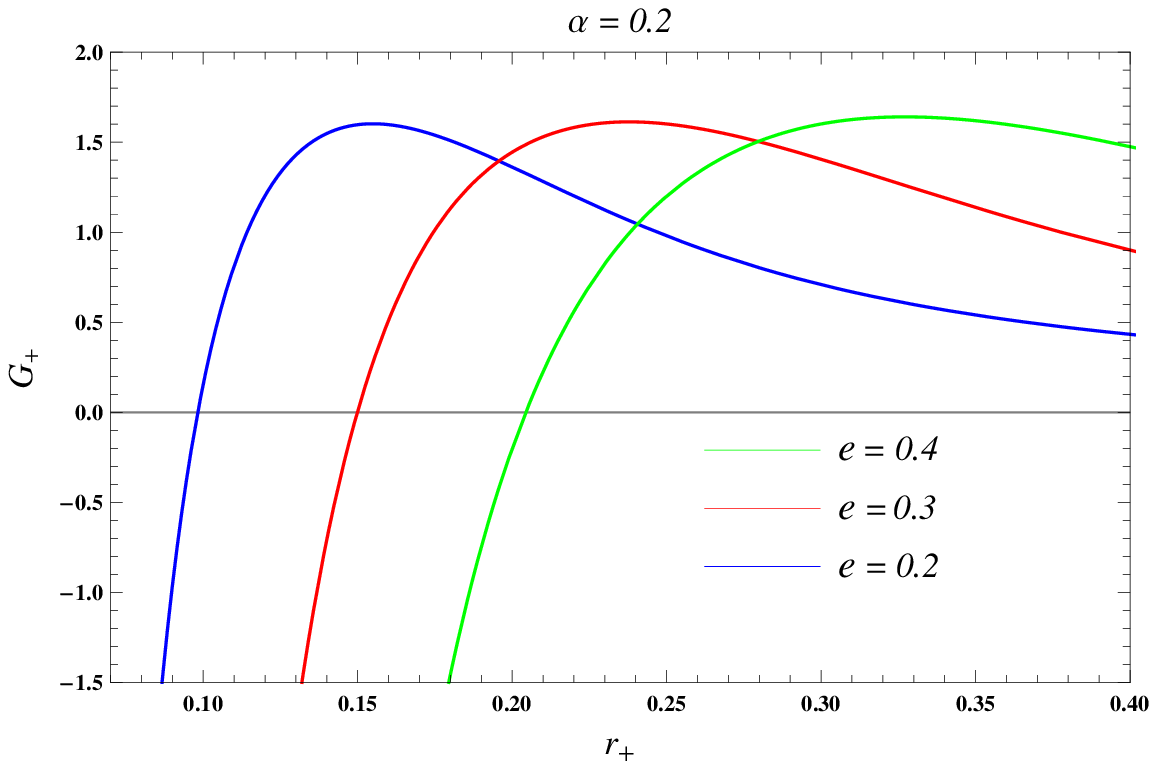}
\end{tabular}
\caption{Free energy $G_+$ \textit{vs} event horizon $r_+$ for various values of $e$ with $\alpha=0.1,\, 0.2$.}
\label{free1}
\end{figure}
The Gibb's free energy for the various value of charge $e$ is shown in Fig. \ref{free1}. The behaviour of Gibb's free energy (cf. Fig. \ref{free1}) dictates the global stability of the black hole. If Gibb's free energy is ($G_+>0$), the black hole is globally unstable, while ($G_+<0$) implies the global stability of the black holes. From Fig. (\ref{free1}), it can be noted that $5D$ EGB-Hayward black holes with smaller event horizon ($r_+$) are globally stable.
The Gibb's free energy will be zero at critical Hawking temperature, so we can calculate critical Hawking temperature $T_+^C$, by using $G_+=0$, as
\begin{equation}
T_C=\frac{3 \left(1+\frac{e^4}{r_+^4}\right) \left(r_+^2+2 \alpha\right )}{4 \pi  r_+ \left[r_+^2+12 \alpha-\frac{e^4}{r_+^4} \left(3 r_+^2+4 \alpha \right)\right]} .
\end{equation}
The black hole will be globally stable when, $T_+>T_+^C$, while $T_+<T_C$, signifies the global instability of the black hole. We have plotted the inverse critical temperature ($T_C$) and inverse temperature ($T_+$) in Fig. (\ref{temp2}), from which we can note the region for which, $T_C^{-1}>T_+^{-1}$, is the region of global stability for our $5D$ EGB-Hayward black hole.

\begin{figure}[ht]
\begin{tabular}{c c c}
\includegraphics[width=0.50\linewidth]{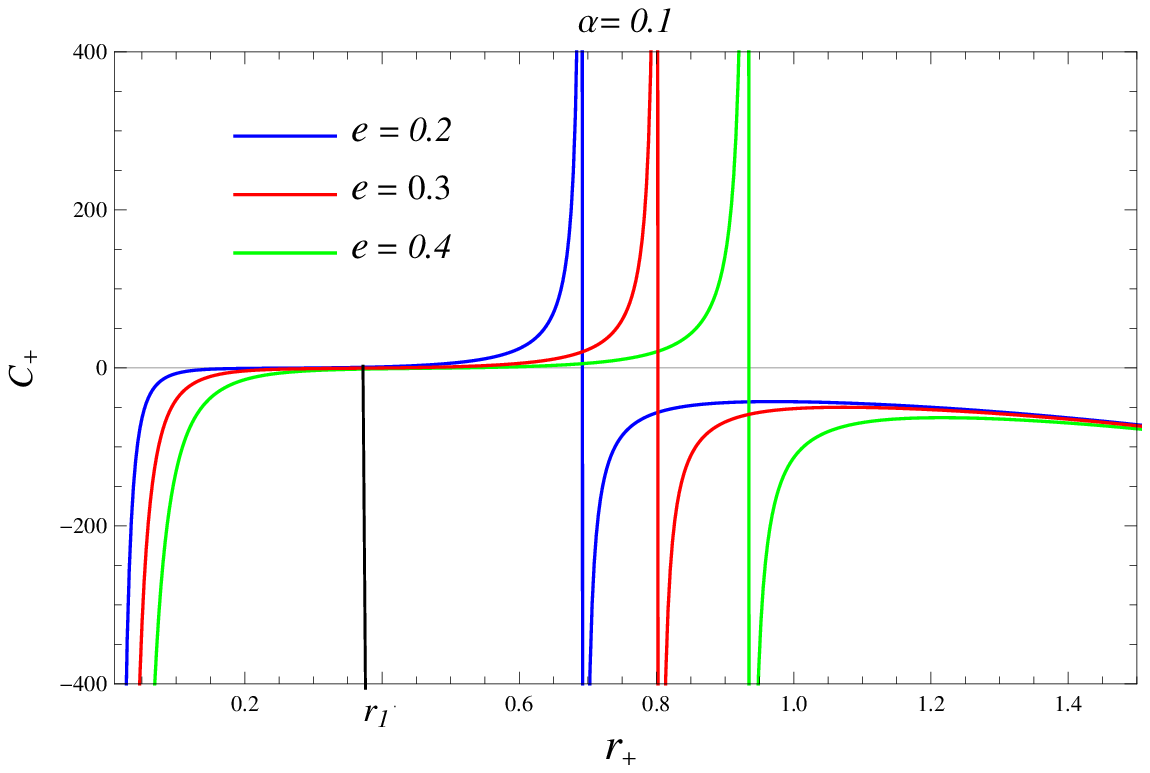}
\includegraphics[width=0.50\linewidth]{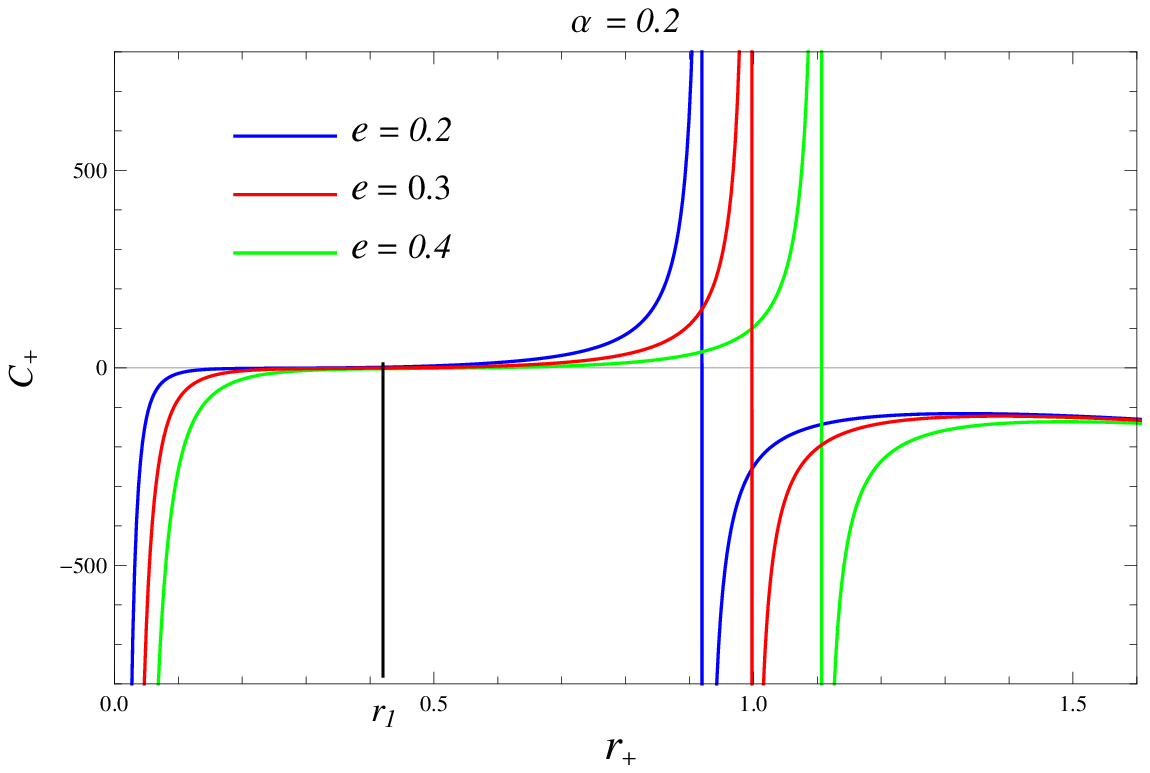}\\

\end{tabular}
\caption{\label{fig:sh} Specific heat $C_+$ \textit{vs} horizon radius $r_+$ for various values of $e$ and $\alpha$.}
\end{figure}

 Thermodynamical stability of $5D$ EGB-Hayward black holes can be performed by studying the behaviour of its specific heat. If the specific heat $C_+>0$, we can say $5D$ EGB-Hayward black holes are thermodynamically stable, while $C_+<0$ indicates that the black holes are thermodynamically unstable. The specific heat of black holes associated with the outer horizon is
 \cite{ ghosh14}
\begin{eqnarray}
C_+=\frac{\partial M_+}{\partial T_+}=\Big(\frac{\partial M_+}{\partial r_+}\Big)\Big(\frac{\partial r_+}{\partial T_+}\Big),
\label{eqC}
\end{eqnarray}
On using Eqs. (\ref{eqM}) and (\ref{eqT}) into Eq. (\ref{eqC}), we get specific heat for $5D$ EGB-Hayward black hole  
\begin{eqnarray}\label{eqc1}
C_+ &=&\frac{24 \pi {V_3}(1+\frac{e^{4}}{r^{4}_+})^2(r^2_++4\alpha)^2r_+}{k_5\left(A(\frac{e^{4}}{r^4_+})^2+B\frac{e^{4}}{r^{4}_+}-Cr^2_+\right)}\left[r_+^2-\frac{e^{4}}{r^{4}_+}(r^2_++4\alpha)\right],
\end{eqnarray}with
\begin{eqnarray}
A=2(r^{2}_++4\alpha)^2, B=16r^4_++40\alpha (3r^2_++4\alpha)\;\;\; \text{and}\;\;\; C=2(r^2_+-4\alpha).\nonumber
\end{eqnarray}
The plots of specific heat with horizon radius $r_+$ are shown in Fig. \ref{fig:sh}. It is well known that the positivity and negativity of the specific heat, respectively, correspond to the thermodynamical stability and instability of the black holes. So, from Fig. (\ref{fig:sh}), one can clearly, notice that our EGB-Hayward black holes are thermodynamically stable in the region $r_1< r_+< r_C$ and unstable for $r_+< r_1$ and $r_+> r_C$.

Eq. (\ref{eqc1}), in the absence of NED, reduced to  
\begin{equation}
C_+=-\frac{{12\pi}{V_3}r_+(r^2_++4\alpha)^2}{k_5(r_+^2-4\alpha)}.
\label{cgb}
\end{equation}
From Eq. (\ref{cgb}), it can be observed that the specific heat diverges as $r_+^2 \rightarrow 4\alpha $, this discontinuity of the specific heat is enough to show the second-order phase transition \cite{hp,davis77} and also the specific heat flips the sign at $ r_+^2=4\alpha $, represents a point of phase transition (Fig. \ref{fig:sh}).
 In the limiting case $\alpha \to0$, one can easily recover the specific heat of $5D$ Hayward black holes can be recovered from Eq. (\ref{eqc1}), in the limit $\alpha \to 0$ 
 \begin{equation}\label{cgb1}
 C_+=\frac{12\pi V_3r_+^3(1+\frac{e^{4}}{r^{4}_+})^2}{k_5(\frac{e^{8}}{r^8_+}+\frac{8e^{4}}{r^{4}_+}-1)}\left[1-\frac{e^{4}}{r^{4}_+}\right], 
 \end{equation}

  which, in the absence of charge NED, further reduces to the specific heat of very well known $5D$ Schwarzschild-Tangherlini black hole \cite{st,ghosh8}.

\subsection{ Black Hole Remnant}
The black hole remnant could play the role of one of the candidates in resolving the information loss puzzle \cite{jp} and also dark matter \cite{jh}. The double root $r_{\pm}=r_E$ of $f(r_E)=0$, represents degenerate horizon of the extremal black hole \cite{Mehdipour:2016vxh,dym1,dym2}.
The remnant radius ($r_E$) of the black hole can be determined from 
\begin{equation}\label{rem}
f'(r_E)=r_E^ˆ6-e^4(r_E^2+4\alpha)=0.
\end{equation}
Eq. (\ref{rem}) can be solved exactly to give remnant radius $r_E$
\begin{equation}
r_E=\sqrt{\frac{3^{\frac{1}{3}}e^4+{\beta_1}^2}{3^{\frac{2}{3}}\beta_1}}~~~~~~\text{with}~~~~~~~\beta_1=\left(18 \alpha  e^4+ \sqrt{3e^8(108 \alpha ^2 -e^{4})}\right)^{\frac{1}{3}}
\end{equation}
 The numerical results of remnants size, minimum mass and maximum temperature for different value of $e$ and $\alpha$ tabulated in Table \ref{tab3}. 
 \begin{center}
\begin{table}[h]
\begin{center}
\begin{tabular}{|l|l r|r l |r l }

\hline
\multicolumn{1}{|c|}{GB coupling }&\multicolumn{1}{c}{ }{\,\,\,\,\,\,$\alpha=0.1$}&\multicolumn{1}{c|}{ }&\multicolumn{1}{c}{ }{\,\,\,\,\,\,\,\,\,\,\,\,\,$\alpha=0.2$ }&\multicolumn{1}{c|}{}\\
\hline
\multicolumn{1}{|c|}{{Charge}}&\multicolumn{1}{c}{ $r_0$ } & \multicolumn{1}{c|}{ $m_0$}&\multicolumn{1}{c}{{$r_0$}}&\multicolumn{1}{c|}{$m_0$}   \\
\hline

\,\,\,\, $e=0.1$\,& \,\,\,\,\,\,\,0.187\,\, &\,\,  0.254\,\, &0.209\,\,\,\,\,\,\,\,\,\,\,&\,\,0.466
\\
\
\,\, $e=0.2$\,\, & \,\,\,\,\,\,\,0.304\,\, &\,\,  0.347\,\,&0.336\,\,\,\,\,\,\,\,\,\,\,&\,\,0.577\,\,
\\
\
\,\, $e=0.3$\,\, & \,\,\,\,\,\,\,0.407\,\, &\,\,  0.473\,\,&0.448\,\,\,\,\,\,\,\,\,\,\,&\,\,0.721\,\,
\\
\
\,\, $e=0.4$\,\, & \,\,\,\,\,\,\,0.506\,\, &\,\,  0.634\,\,&0.552\,\,\,\,\,\,\,\,\,\,\,&\,\,0.899\,\,
\\
\
\,\, $e=0.5$\,\, & \,\,\,\,\,\,\,0.602\,\, &\,\,  0.830\,\,&0.651\,\,\,\,\,\,\,\,\,\,\,&\,\,1.110\,\,
\\
 \hline
\end{tabular}
\end{center}
\caption{The remnant size $r_0$ and remnant mass $m_0$ for different values of  charge $e$.}
\label{tab3}
\end{table}
\end{center}
\begin{figure}[ht]
\begin{tabular}{c c c c}
\includegraphics[width=0.5\linewidth]{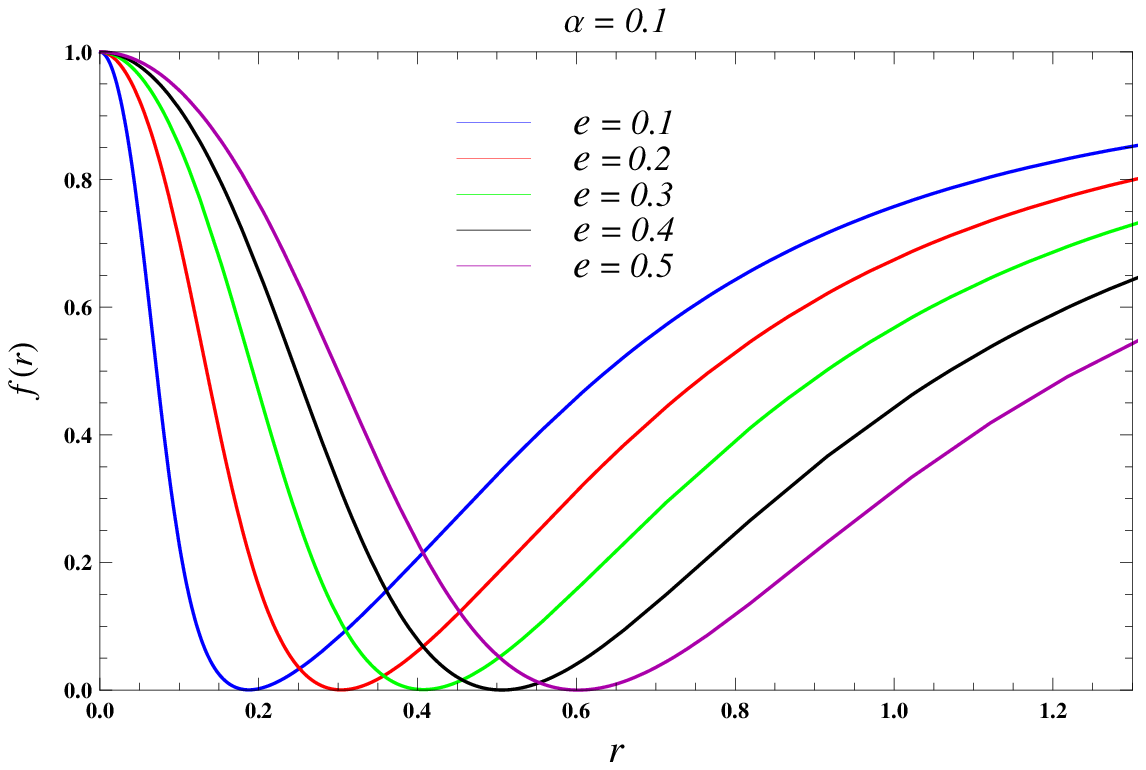}
\includegraphics[width=0.5\linewidth]{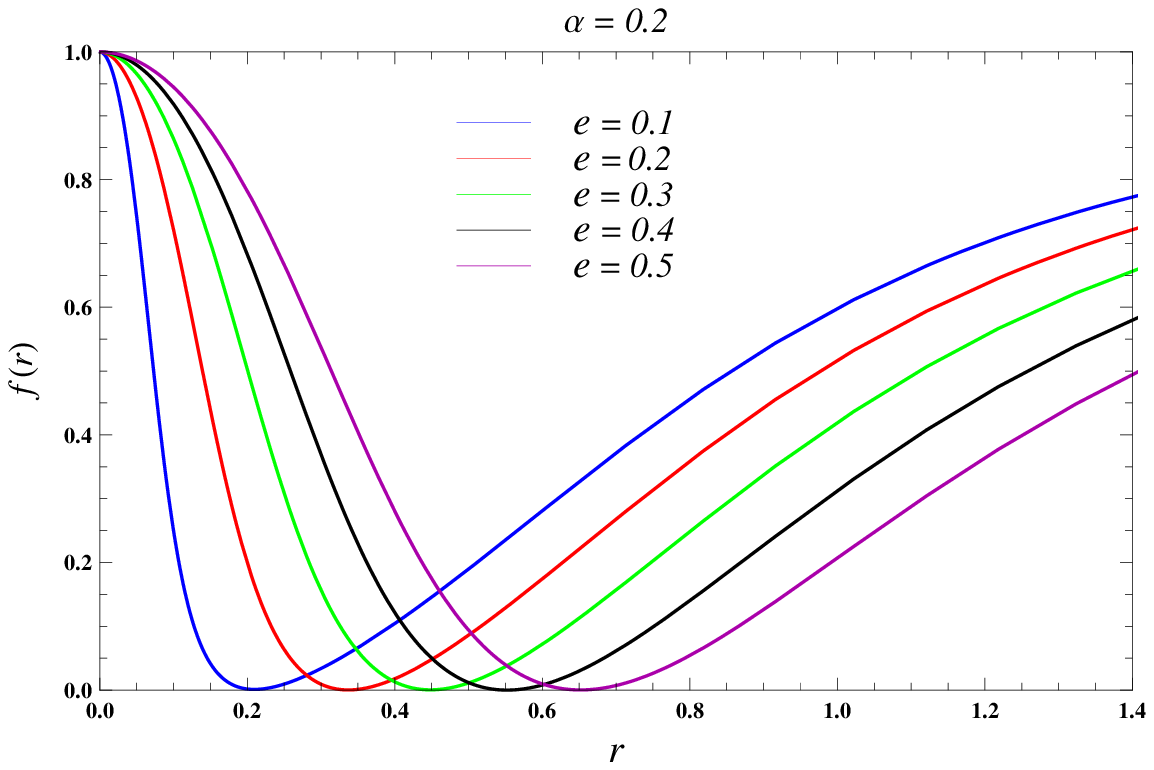}
\end{tabular}
\caption{ Metric function $f(r)$ \textit{vs} horizon radius $r_+$ with various values of $e$ and $\alpha$.}
\label{rem1}
\end{figure}
\begin{figure}[ht]
\begin{tabular}{c c c c}
\includegraphics[width=0.5\linewidth]{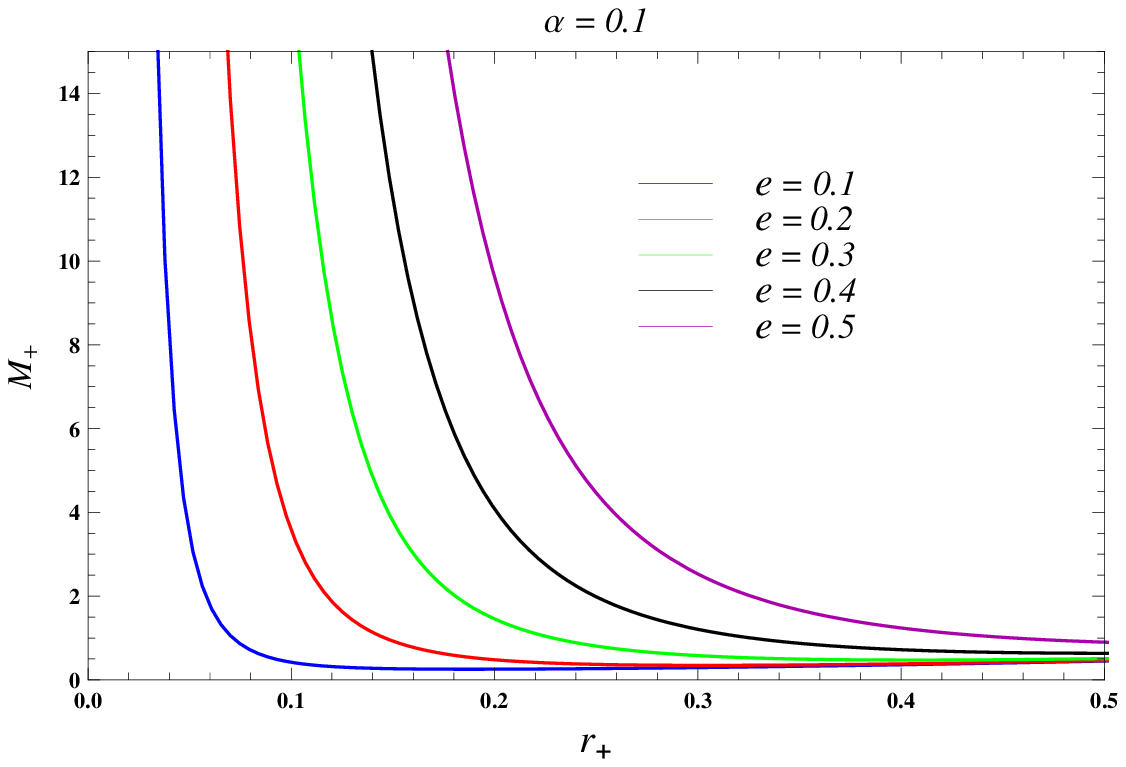}
\includegraphics[width=0.5\linewidth]{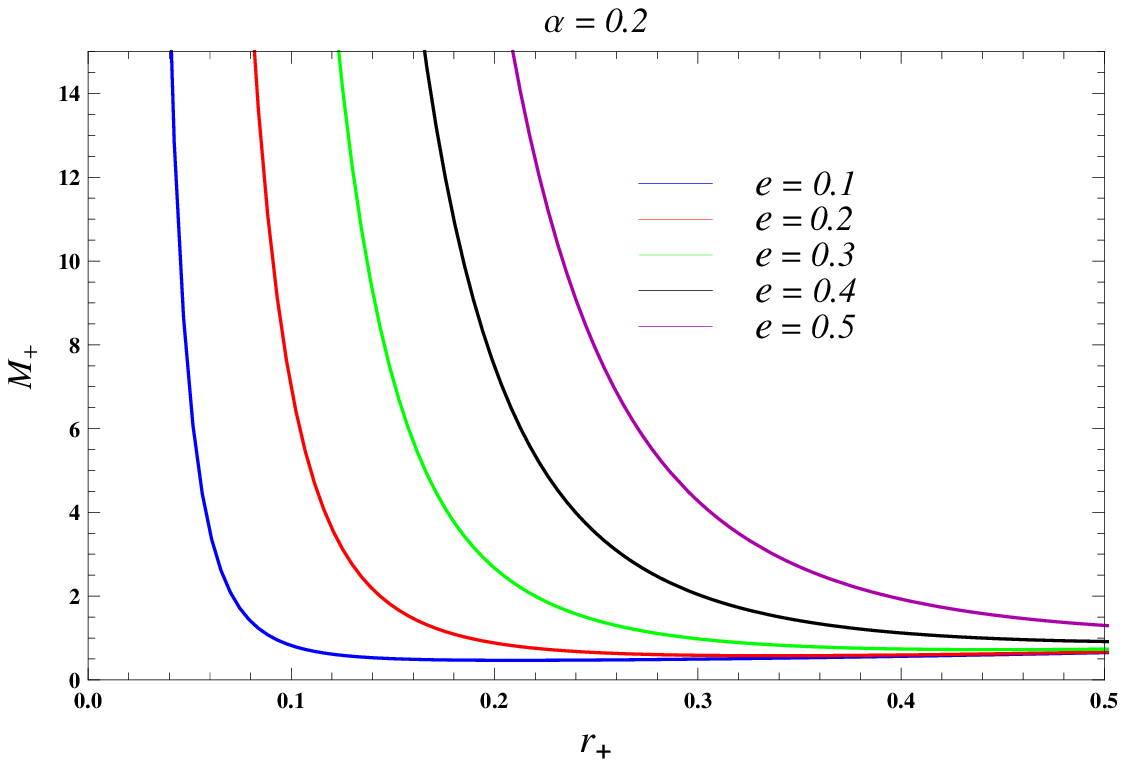}
\end{tabular}
\caption{Black hole mass $M_+$ \textit{vs} event horizon ($r_+$) with various values of $e$.}
\label{rem2}
\end{figure}
 The temporal component of the metric, $f(r)$, of $5D$ EGB-Hayward black hole with horizon radius for different values of $e$ has been displayed in Fig.\ref{rem1}. From Fig.\ref{rem1}, we can notice that an extremal configuration with one degenerate event horizon at a minimal nonzero mass $m_0$ is possible. Hence, the EGB-Hayward black hole can evaporate to leave a stable remnant of mass $m_0$. In fact, the condition for having one degenerate event horizon is that $m = m_0$ which means for $m < m_0$ there is no event horizon (cf. Fig. \ref{rem1} and Fig. \ref{rem2}). In the near extremal region temperature increase from zero to local maximum $T_+^{Max}$ corresponding to $r_+=r_{max}$. As $r_+$ increase further $T_+$ drops to local minimum corresponding to $r_+=r_{min}$.

\section{Conclusion}
We have obtained an exact Hayward-like regular black holes in EGB coupled to nonlinear electrodynamics. The solution has an additional parameter charge $e$ due to nonlinear electrodynamics, apart from the black hole mass ($M$). The previously known case like the famous Boulware-Desser black holes of EGB theory is encompassed as a special case. In turn,  we characterized the solution by analyzing horizons which are maximum three, viz. Cauchy, Event and cosmological horizons. The regularity of the spacetime is confirmed by calculating various curvature invariants and shown to be well behaved everywhere including at origin. We have also computed thermodynamical quantities like Hawking temperature, entropy, specific heat and free energy associated with $5D$ EGB-Hayward black holes with a focus on the stability of the system. It is demonstrated that specific heat diverges at horizon radius $r_+^c$ which incidentally corresponds to the local maximum of the Hawking temperature at $r_+^c$. It is shown that the specific heat is positive in the region $r_1<r<r_+^c$, which signifies that the small black holes are thermodynamically stable against perturbations in the region, and the phase transition exists at $r_+^c$. The black hole is unstable for $r_1>r>r_+^c$. The global stability analysis of black holes is also done by calculating free energy. Besides, we have also shown that after black hole evaporation there will be a stable remnant with zero temperature and positive specific heat. 
Further,  a generalization of such a  regular black hole configuration to Lovelock gravity is an important direction for the future. Also, it would be interesting to generalize this solution by including the AdS background.

\begin{acknowledgements}
Authors would like to thank IUCAA for the hospitality while this work was being done.
\end{acknowledgements}

\end{document}